\documentclass[reprint,amsmath,amssymb,aps,pre]{revtex4-1}

\usepackage{graphicx}  
\usepackage{dcolumn}   
\usepackage{bm}        
\usepackage{amssymb}   
\usepackage{mathtools}
\usepackage{amsmath}
\usepackage{color}
\usepackage{dblfloatfix}

\begin{document}

\title{Learning by non-interfering feedback chemical signaling\\in physical networks}
\author{Vidyesh Rao Anisetti,$^1$ B. Scellier,$^{2}$ J. M. Schwarz$^{1,3}$}

\affiliation{$^1$ Physics Department, Syracuse University, Syracuse, NY 13244 USA\\
$^2$ Department of Mathematics, ETH Z\"{u}rich, Z\"{u}rich, Switzerland\\
$^3$ Indian Creek Farm, Ithaca, NY 14850 USA }

\date{\today}
\begin{abstract}
Both non-neural and neural biological systems can learn. So rather than focusing on purely brain-like learning, efforts are underway to study learning in physical systems. Such efforts include equilibrium propagation (EP) and coupled learning (CL), which require storage of two different states---the free state and the perturbed state---during the learning process to retain information about gradients. Here, we propose a new learning algorithm rooted in chemical signaling that does not require storage of two different states. Rather, the output error information is encoded in a chemical signal that diffuses into the network in a similar way as the activation/feedforward signal. The steady state feedback chemical concentration, along with the activation signal, stores the required gradient information locally. We apply our algorithm using a physical, linear flow network and test it using the Iris data set with 93\% accuracy. We also prove that our algorithm performs gradient descent. Finally, in addition to comparing our algorithm directly with EP and CL, we address the biological plausibility of the algorithm.

\end{abstract}
\maketitle

\section{Introduction}
What basic ingredients constitute a biological learning system, such as slime mold or higher-order organisms? Biological learning systems adapt to the external environment by tailoring specific responses for given external conditions. As the system continues to experience external conditions of a similar kind, it develops functionality to respond to the stimulus in such a way to increase its chances of survival. Intriguingly, this functionality is an emergent phenomenon as a result of interactions between the various components~\cite{Sumpter2006}. For example, when birds come together in a flock, they increase their chances of survival~\cite{Beauchamp2021}. This happens not because of a ‘supervisor’ that commands each bird to fly in a particular way, but because birds, such as starlings, interact with a fixed number of neighbors independent of their density to give rise to emergent functionality~\cite{Ballerini2008}. Similarly, in the presence of rising waters, fire ants cooperate to form floating rafts consisting of a structural base and freely-moving ants on top of the base with treadmilling between the two roles~\cite{antraft1,antraft2}. Local, ant interaction rules, including an effective repulsive force between the freely-moving ants and the water replicate the types of observed shapes of rafts~\cite{wagner2022}. These special interactions between components, responsible for emergent functionality in nature, have themselves emerged out of the long process of evolution.

Given the intricacies of biological learning systems, neural networks are {\it in silico} brain-like learning systems, resembling the visual cortex, in particular~\cite{Issa2018,Kriegeskorte2015,D1988,Roelfsema2018}, that can recognize patterns and solve problems~\cite{nielsenneural,Yang2020}. More specifically, neural networks achieve functionality by modifying weights and biases to minimise a particular cost function. Of the many ways to do so, the algorithm of choice in neural networks with multiple layers (deep learning) is the backpropagation algorithm~\cite{Rumelhart1986}. Backpropagation updates the network such that its weights (and biases) perform gradient descent in the cost function landscape. The complex nature of the tasks that neural networks are capable of hints at the possibility that biological learning systems also achieve functionality by optimizing cost functions by gradient descent~\cite{Lillicrap2020}. In other words, the long process of evolution may have optimised the ``learning algorithm" in such biological systems to update its components via gradient descent. The success of backpropagation has, therefore, encouraged a search for biologically plausible learning rules analogous to it~\cite{sacramento2018dendritic,Pozzi2018,Richards2019,WHITTINGTON2019235,Lillicrap2020,payeur2021burst}. For completeness, here are properties one should ensure while constructing such a biologically plausible learning system :
\begin{enumerate}
    \item local learning algorithms~\footnote{By local we not only mean in terms of metric distance},
    \item the implementation of such algorithms is constrained by the laws of physics, and 
    \item the algorithms minimize a cost function via gradient descent or stochastic gradient descent.
\end{enumerate}

Indeed, there have been attempts to construct learning algorithms within purely physical systems~\cite{Hopfield1984,markovic2020physics,endtoend2020,scellier2020deep,Stern2021,wright2022deep}. Here, we will focus on ``equilibrium propagation"~\cite{Scellier2017,scellier2020deep} and ``coupled learning"\cite{Stern2021}. In these approaches, the error information corresponding to each component is encoded in terms of differences of local physical quantities measured between two phases. At each step of training, the outputs of the network are nudged towards the target output by applying additional boundary conditions at the output nodes. Next, the system is allowed to settle to a new steady state called the  ``nudged state" (or the ``clamped state"), which is closer to the desired target than the initial ``free state". In the limit where the nudge amplitude goes to zero, the difference of local, physical quantities between these two phases encodes the gradient of the cost function~\cite{Scellier2017}. Unlike backpropagation, these algorithms achieve gradient descent without an explicit layer-by-layer transfer of error information. Nevertheless, one caveat of these approaches is the requirement to store the free state. In other words, the learning rule requires information about the free state, which is no longer physically available at the end of the second phase when the parameters are updated.  One way around this requirement is to build two copies of the same physical network~\cite{Dillavou2021}, though biology does not necessarily have such a luxury.

In this work, we present a new algorithm and a new learning rule that overcomes the above requirement in equilibrium propagation and coupled learning.
Our learning rule computes the gradients using local information for each weight without the need to store the free state. We demonstrate that the functionality of the nudged state can be realised in physical and biophysical, learning systems using chemical signalling. We show that steady state chemical concentrations can be used in the second phase to encode the required gradient information. Chemical signaling is ubiquitious in biology.  For example, consider the structurally simple, yet functionally complex organism, named {\it Physarum polycephalum}, otherwise known as slime mold. Slime mold is a uni-cellular, multi-nucleated organism that is neither a plant nor an animal nor a fungus. This uni-cellular organism can span up to the meter scale  and consists of a network of tubes whose underlying structure is driven by cytoskeletal reorganization~\cite{slimemold}. Despite its simplicity, in the sense that it is non-neuronal, this organism is capable of myriad complex tasks -- precisely coordinating flows in its body~\cite{Boussard2021}, navigating mazes~\cite{slimemaze2000}, and connecting food sources with optimal paths~\cite{tero2010,Tero2008}. Work by K. Alim and others showed that much of this complex phenomenon can be explained by a mechanism of signal propagation~\cite{Alim2017}. Specifically, slime mold uses a chemical signal to send information regarding the location of food sources across its body, which triggers a change in its tubular structure due to a softening agent to optimize the connection between food sources~\cite{Alim2021}.\\ \\

In light of an example, we construct a physical learning network of tubes/pipes that uses chemical signals to send error information across the system. Our system is a flow network, with activation pressures at nodes \textbf{v} and pipe conductance described by weights \textbf{w}. The information from the external environment is input into the system by fixing the boundary conditions \textbf{I} at input nodes. Node activation pressures \textbf{v} are the non-trainable variables (or ``state variables'') of the system that are determined by Laplace's equation and input boundary conditions. Our physical system is, therefore, a linear one. The functionality we seek is to obtain desired pressures (``target pressures") at output nodes for a given input \textbf{I}. To achieve this, a feedback chemical is released into the flow network by fixing the chemical currents at output nodes. The value $\boldsymbol{\epsilon}$ of these chemical currents is proportional to the difference between target pressures and output pressures. The chemical concentrations \textbf{u} at internal nodes are determined by the same Laplace equation, but with feedback boundary conditions $\boldsymbol{\epsilon}$.  We show that this chemical concentration \textbf{u} along with the node pressures \textbf{v} locally encode the weight gradients of the cost function that we want to optimize. We propose a new learning rule that updates the trainable weights \textbf{w} such that it does gradient descent with respect to the cost function.

\section{Theory}
\label{sec:theory}

\begin{figure}[t]
    \centering
    \includegraphics[width=8cm]{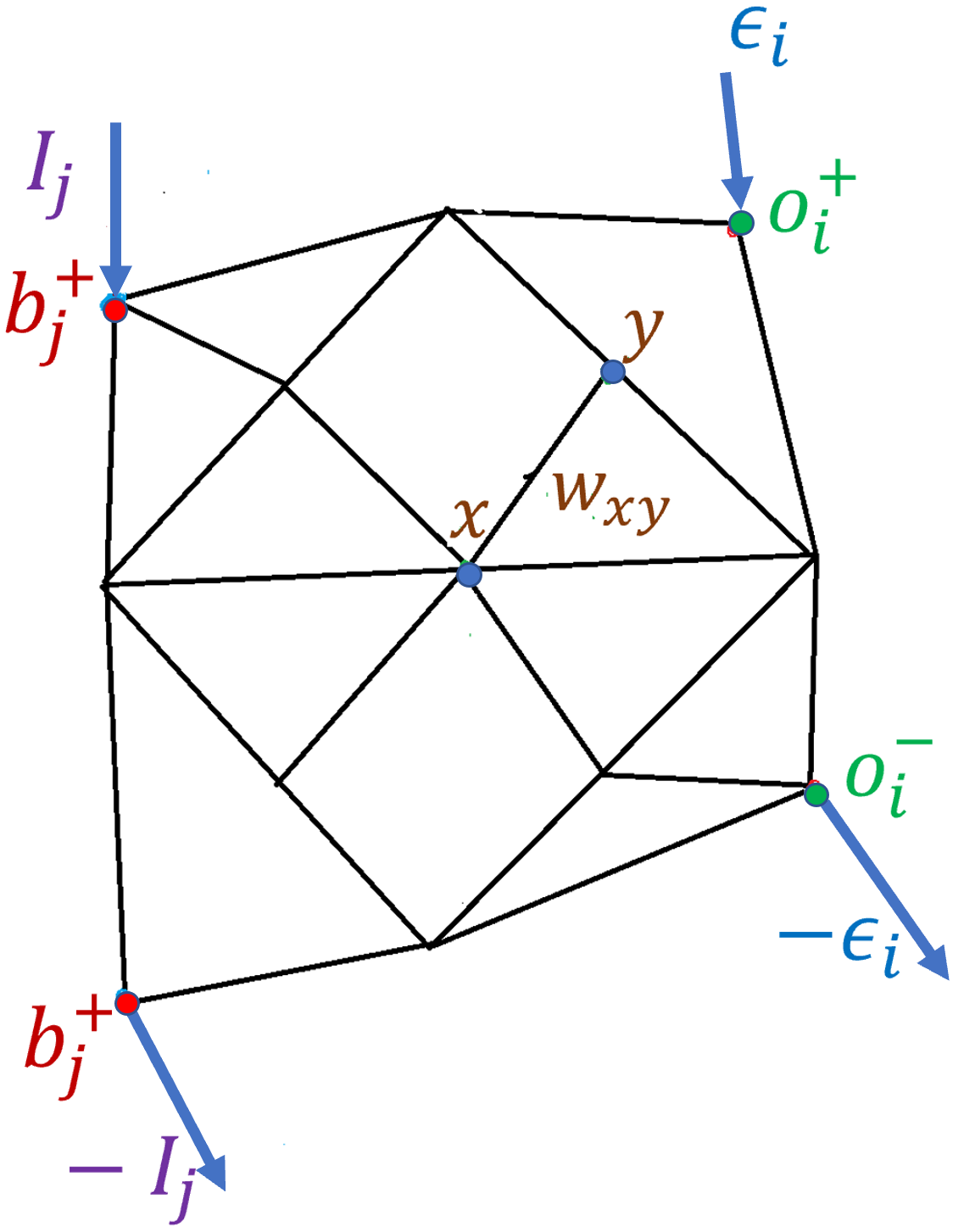}
    \caption{{\it The flow network.} Schematic of the flow network with blue points representing the input node pair and red points representing the output node pair. The weights of the flow network are denoted by $w_{xy}$ between neighboring nodes $x$ and $y$. These weights are varied during the training-testing process.}   
    \label{fig:1}
\end{figure}

We consider a flow network of nodes interconnected by weighted edges. We denote $w_{xy}$ the weight (i.e. conductance) of the edge between node $x$ and node $y$. A subset of the nodes are boundary node pairs (or ``input'' node pairs), denoted $\{(b_1^+,b_1^-), (b_2^+,b_2^-), \ldots, (b_q^+,b_q^-)\}$. For each pair $(b_j^+,b_j^-)$, an input current $I_j$ flows into the network through the node $b_j^+$ and flows out of the network through the node $b_j^-$. The state of the system is defined by the node pressures, denoted $v(x)$ and governed by Laplace's equation at steady state. Another subset of the nodes are output node pairs $\{(o_1^+,o_1^-), (o_2^+,o_2^-), \ldots, (o_p^+,o_p^-)\}$.
The output of the network is defined as the set of pressure drops across output nodes $\{v(o_1^+,o_1^-), v(o_2^+,o_2^-), \ldots, v(o_p^+,o_p^-)\}$ where $v(o_i^+,o_i^-) = v(o_i^+)-v(o_i^-)$. We note that $v(o_i^+,o_i^-)$ is a function of input currents $\{I_j\}$ and all the weights of the network $\{w_{xy}\}$. Training the network consists in modifying the weights $\{w_{xy}\}$ such that, given the input currents $\{I_j\}$, we get desired pressure drops $\{v_d(o_i^+,o_i^-)\}$ across the output node pairs. We define the cost function
\begin{equation}
    \label{eq:cost-function}
    C = \frac{1}{2} \sum_{i=1}^{p} (v(o_i^+,o_i^-)-v_d(o_i^+,o_i^-))^2.
\end{equation}
Now we present a physical procedure and a learning rule for the weights that performs gradient descent with respect to the cost function. To achieve this, we release a feedback chemical into the network through the pairs of output nodes $\{(o_i^+,o_i^-)\}$, by fixing chemical currents across these nodes. Specifically, for each pair of output nodes $(o_i^+,o_i^-)$, a current
\begin{equation}
    \label{eq:chemical-current}
    \epsilon_i = \eta (v_d(o_i^+,o_i^-)-v(o_i^+,o_i^-))
\end{equation}
flows into the network through node $o_i^+$ and flows out of the network through node $o_i^-$, where $\eta$ is a constant (``nudging''). A steady state chemical concentration develops at every node, governed again by Laplace's equation. We denote $u(x)$ the steady state concentration at node $x$, and $u(x,y)$ the drop in chemical concentration between nodes $x$ and $y$. Finally, we update each weight $w_{xy}$ according to
\begin{equation}
    \label{eq:learning-rule}
    \Delta w_{xy} = - \alpha u(x,y)v(x,y), 
\end{equation}
where $\alpha$ is a constant. We show below that this learning rule performs gradient descent on the cost function with step size (``learning rate'') $\alpha \eta$, i.e.
\begin{equation}
    \label{eq:gradient-descent}
    \Delta w_{xy} = - \alpha \eta \dfrac{\partial C}{\partial w_{xy}}
\end{equation}
for every weight $w_{xy}$. The above learning rule for the weights is local. The final error term depends upon two quantities; the pressure drop due to flow and the concentration drop of the feedback chemical. If they have the same sign, then the weight gets a positive update and vice versa. Here, we assume that the relaxation time scale of the system is much faster than the time scale of weight updates so that the system is in steady state as the weights are updated. 

Note that the two quantities in the weight update are independent of each other, therefore, we assume that diffusion is fast enough that it is independent of the flow. In an experimental setting, one can realize this by using very small flow rate, leading to very small pressure drops across weights and using a signalling chemical with very high diffusion rate via, perhaps, some catalytic process. We understand that such a construction is not necessarily realized in nature, therefore, we also propose a purely flow version of the model where the chemical signals are carried by the fluid flow and not by diffusion (Appendix A). While this chemical flow algorithm is presumably more physically plausible, it is not yet clear that the algorithm performs gradient descent. In any event, what we present here is an idealization. Obviously, nature may be using a complex combination of flow and diffusion for signalling. 

 Considering $v$ as a pressure and $u$ as a chemical concentration is just a certain packaging of the theory. The central idea of this work is to use two independent physical quantities which leads to non-interference of the input signal and the feedback signal, in other words one can use any two non-interfering modalities to conduct learning~\footnote{ We acknowledge Nachi Stern for suggesting the use of the word \textit{modality} }.  For example, one can use two chemicals $v$ and $u$ diffusing in a static fluid, with distinct chemical signatures to encode input and error signal. In this case, there is no need for an additional assumption on the relationship between flow rate and diffusion rate. 

Our result holds for any cost function $C$, not just the squared error (Eq.~\ref{eq:cost-function}). In general, in the second phase, the chemical current flowing in through $o_i^+$ and flowing out through $o_i^-$ must be $\epsilon_i = - \eta \frac{\partial C}{\partial v(o_i^+,o_i^-)}$. Our result also holds if we reverse the sign of $\eta$ \eqref{eq:chemical-current} and the sign of $\alpha$ in the learning rule \eqref{eq:learning-rule}. In other words, the algorithm performs gradient descent so long as $\alpha\eta>0$. 



Now we prove our claim that the learning rule \eqref{eq:learning-rule} performs gradient descent \eqref{eq:gradient-descent}.
Let us number the nodes of the network $1,2,\ldots,n$. Let $I_x$ be the input current at node $x$ (with $I_x = 0$ by convention if node $x$ is not an input node) and $v_x$ the pressure at node $x$. For each node $x$, the steady state condition at node $x$ in the first phase yields:
\begin{equation}
    \sum_y w_{xy} (v_x-v_y) = I_x.
\end{equation}
We get a system of $n$ linear equations. This system rewrites with matrix-vector notations as
\begin{equation}
    \label{eq:first-phase}
    L \cdot v = I,
\end{equation}
where $v$ is the vector of node pressures, $I$ is the vector of input currents and $L$ is the matrix
\begin{equation}
\label{eq:matrix-L}
L = 
\left( \begin{array}{ccccc}
\sum_{x} w_{1x}     & -w_{12} & -w_{13} & \cdots & -w_{1n} \\
-w_{21} & \sum_{x} w_{2x}   & -w_{23} & \cdots & -w_{2n} \\
-w_{31} & -w_{32} & \sum_{x} w_{3x}   & \cdots & -w_{3n} \\
\vdots  & \vdots  & \vdots  & \ddots & \vdots \\
-w_{n1} & -w_{n2} & -w_{n3} & \cdots & \sum_{x} w_{nx} \\
\end{array} \right).
\end{equation}
Note that the matrix $L$ is symmetric because $w_{xy} = w_{yx}$ for every pair of nodes $(x,y)$.
Next, we denote $E_x=-\eta \frac{\partial C}{\partial v_x}$ the chemical current at node $x$ in the second phase (with $e_x=0$ by convention if node $x$ is not an output node), and $u_x$ the concentration of the chemical at node $x$. Assuming that the diffusion constant of the chemical is equal to the flow conductivity (up to a constant of proportionality), the chemical concentration at steady state satisfies the same system of linear equations, with different boundary conditions, i.e.
\begin{equation}
    \label{eq:second-phase}
    L \cdot u = E,
\end{equation}
where $u$ is the vector of node concentrations, and $E$ is the vector of chemical currents.
Now we compute the gradient of the cost function: for every weight $w_{xy}$ we have
\begin{equation}
    \frac{\partial C}{\partial w_{xy}} = \left( \frac{\partial C}{\partial v} \right)^\top \cdot \frac{\partial v}{\partial w_{xy}}.
\end{equation}
Multiplying by $-\eta$ on both sides, we get, by definition of $E$,
\begin{equation}
    -\eta \frac{\partial C}{\partial w_{xy}} = E^\top \cdot \frac{\partial v}{\partial w_{xy}}.
\end{equation}
Using the steady state condition of the second phase \eqref{eq:second-phase} and the fact that $L$ is symmetric, we get
\begin{equation}
    \label{eq:proof-1}
    -\eta \frac{\partial C}{\partial w_{xy}} = u^\top \cdot L \cdot \frac{\partial v}{\partial w_{xy}}.
\end{equation}
Next, we differentiate the steady state condition of the first phase \eqref{eq:first-phase} with respect to $w_{xy}$. We get
\begin{equation}
    \frac{\partial L}{\partial w_{xy}} \cdot v + L \cdot \frac{\partial v}{\partial w_{xy}} = 0.
\end{equation}
Rearranging the terms and injecting this in \eqref{eq:proof-1}, we get
\begin{equation}
    \label{eq:proof-2}
    \eta \frac{\partial C}{\partial w_{xy}} = u^\top \cdot \frac{\partial L}{\partial w_{xy}} \cdot v.
\end{equation}
Looking at the form of the matrix $L$ \eqref{eq:matrix-L}, the matrix $\frac{\partial L}{\partial w_{xy}}$ has exactly four non-zero coefficients: those at positions $(x,x)$, $(x,y)$, $(y,x)$ and $(y,y)$, equal to $+1$, $-1$, $-1$ and $+1$, respectively. Therefore
\begin{align}
    \nonumber
    & u^\top \cdot \frac{\partial L}{\partial w_{xy}} \cdot v \\
    = \;\; & u(x)v(x)-u(x)v(y)-u(y)v(x)+u(y)v(y) \\
    = \;\; & (u(x)-u(y)) \cdot (v(x)-v(y)) \\
    = \;\; & u(x,y) \cdot v(x,y).
\end{align}
Combining this with \eqref{eq:proof-2}, we conclude that
\begin{equation}
    \Delta w_{xy} = - \alpha u(x,y) \cdot v(x,y) = -\alpha \eta \frac{\partial C}{\partial w_{xy}}.
\end{equation}
Hence, the learning rule corresponds to one step of gradient descent with step size $\alpha \eta$. This concludes the proof.

Summing up the training mechanism:
\begin{enumerate}
    \item A flow network is generated where chemicals can spread via a diffusion process. 
    \item Input to the network is given by fixing currents $\{I_j\}$ across input node pairs $\{(b_j^+,b_j^-)\}$, leading to steady state pressures $v$. Outputs are measured as pressure drops $\{v(o_i^+,o_i^-)\}$ across output node pairs $\{(o_i^+,o_i^-)\}$.
    \item Outputs are compared to the desired outputs $\{v_d(o_i^+,o_i^-)\}$, and a feedback chemical is released in the network by fixing the chemical currents $\epsilon_i = \eta (v_d(o_i^+,o_i^-)-v(o_i^+,o_i^-))$ across the output node pairs $\{(o_i^+,o_i^-)\}$. The chemical concentration reaches steady state $u$.
    \item The concentration drop $u(x,y)$ and pressure drop $v(x,y)$ are measured across each weight $w_{xy}$, and the weights are updated according to $\Delta w_{xy} =  - \alpha u(x,y)v(x,y)$.
    \item This procedure, which corresponds to one step of gradient descent, is repeated iteratively until convergence of the weights is achieved.
\end{enumerate}

\section{The Iris data set}

We train the flow network on a standard machine learning task: classifying Iris flowers. The Iris data set\cite{misc_iris_53} contains 150 examples of Iris flowers belonging to three species (setosa, virginica and versicolor), and, therefore, 50 examples for each category. Each example is of the form $X = (X_1, X_2, X_3, X_4)$, composed of four features of the flower (petal width, petal length, sepal width and sepal length, all measured in cm), and comes with its assigned iris category, denoted $Y$. So an example would look something like $X = (5.1,3.5,1.4,0.2)$ and $Y=\text{`setosa'}$. Given the four features of the flower as input, the trained network should be able to tell which species it belongs to. 
\begin{figure*}[t]
    \centering
    \includegraphics[width=16cm]{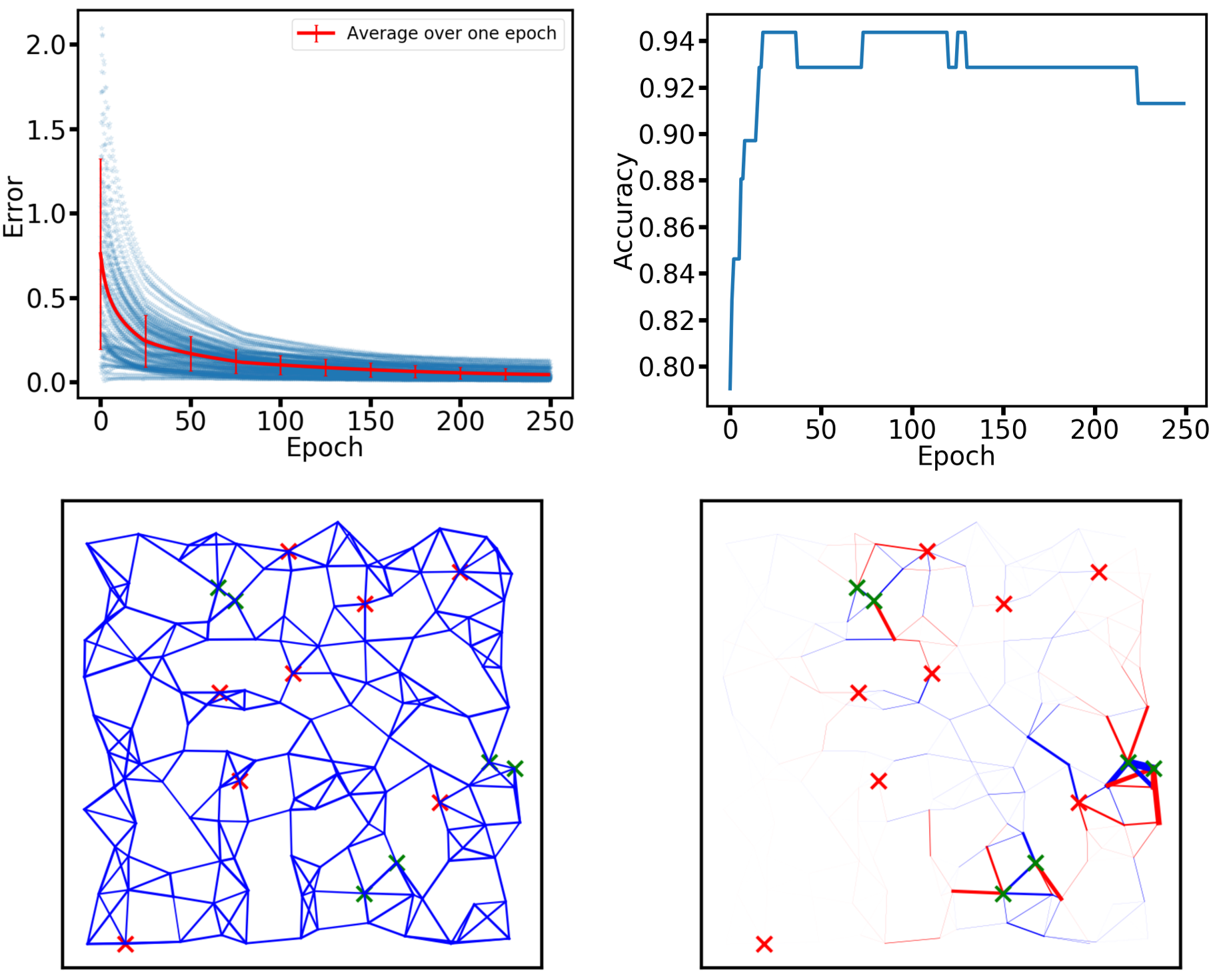}
    \caption{{\it Training and testing using the Iris data set.} The Iris data set is trained on a network with $12^2$ nodes and  learning rate $\eta = 10^{-4}$. The weights are sampled from a truncated normal distribution with mean of 0.1, standard deviation of 0.01 with a lower cutoff of 0.01 and upper cutoff of 0.19. We use the standard mean squared error (L2 norm) as the cost function. [Top Left] At each training step, the network is shown an example and the mean squared error is calculated between the network's output and desired output, shown as a blue dot. [Top Right] Accuracy is defined as the fraction of correct predictions out of total testing examples.  [Bottom Left] The initial state of the network, with thickness of blue edges representing the conductance. Red crosses denote input nodes; green crosses denote output nodes. [Bottom Right] The change in weights between the initial state and the final trained state of the network. Red denotes negative change, while blue denotes positive change with thickness showing the magnitude of change.
    }   
    \label{fig:2}
\end{figure*}

We now detail how we do this. A flow network is constructed as follows:
    \begin{enumerate}
        \item Generate a square lattice,
        \item Perturb the positions of the lattice with a step of length $\delta$ in any random direction,
        \item every node is connected to its $d$ nearest neighbours \footnote{Note that $d$ may not be equal to degree of the node: given two neighbouring nodes $A$ and $B$, it is possible that $B$ is in the $d$ nearest neighbours of $A$, while $A$ is \textit{not} in the $d$ nearest neighbors of $B$.}. We choose $d = 4$ for all our simulations unless specified.
        \item Every edge of the network is assigned a conductance from a truncated normal distribution. 
         \item Four pairs of input nodes are chosen from this flow network, where the input data (the normalised features of the iris) is given as external currents across these four pairs of boundary nodes. Three pairs of nodes are chosen as the output nodes such that every pair is composed of two neighbouring nodes. For pairs of output nodes that are not neighboring nodes, the training error decreased less smoothly. Once the network is trained, for a given input, the set of potential drops across these node pairs should tell the category of iris the input data corresponds to. 
    \end{enumerate}        
The network architecture remains fixed throughout the training-testing process. Only the conductances of these weights are modified.   

As for how the flow network interfaces with the Iris data set, 
\begin{enumerate}
    \item The data set is divided into two subsets: one training set (used for training) and one test set (used for testing). Each of these sets have 75 examples of irises, 25 from each category.
    \item The data set is normalised. That is, for each example $X$ in the data set, and for each feature $X_i$ of $X$; we set $X_i^{\rm norm} = \lambda \cdot \dfrac{X_i-X_i^{\rm min}}{X_i^{\rm max}-X_i^{\rm min}}$, where $X_i^{\rm max}$ and $X_i^{\rm min}$ are the minimum and maximum values for that feature $X_i$ in the training set. We choose $\lambda=5$ for all simulations.
\end{enumerate}

While choosing the desired outputs we must keep in mind the fact that this linear system may not be able to find a set of weights that give out the desired output. In other words, we must choose desired outputs that are physically attainable. We, therefore, implement the same technique as described in Ref.~\cite{Dillavou2021} to choose the desired output voltages for each of the three Iris categories. For each category, the desired voltage is the average, normalized input data. That is, each iris category has 25 examples of four input features, each input feature is averaged out over 25 examples. Finally, we obtain a four tuple of averaged input features for each iris category. When this is given as input to the initial network, we aim to arrive at an output voltage that corresponds to the average behavior of the input, which is the  desired voltage.

To conduct the training process, first the input data is given to the network and the output is observed. If the output is not equal to the desired output for that iris category, the feedback chemical is released at the output node pairs by fixing the currents. The weights of the network are modified using the learning rule mentioned above. This process is repeated consecutively for all examples. Next, once the entire data set is exhausted, we say 'one epoch' has passed. We train the network for multiple epochs. At the beginning of each epoch, because the network has changed significantly, new desired voltages are calculated. Therefore, each epoch has its own set of desired voltages.

To conduct the testing process, after the network is trained for multiple epochs, we record how well it classifies unseen data from the test set. To be specific, the test set has 25 examples per iris category. The iris features from these test examples are given as input and the output voltage is compared with the desired voltage. The desired voltage is calculated using the testing data set. An example is then classified into that iris category, for which the output voltage is closest to the desired output.

This training-testing procedure is implemented on a biophysical network. Results are shown in Fig 2.

\section{Link with Equilibrium Propagation and Coupled Learning}
\label{sec:related-work}

To readily see the link with our algorithm and EP,
suppose that, in the second phase, instead of injecting the chemical, we inject more of the main substance at output nodes (through the currents $\epsilon_i = \eta (v_d(o_i^+,o_i^-) - v(o_i^+,o_i^-))$).
We can adapt the analysis of Sec.~\ref{sec:theory} to show that, instead of the chemical concentration, $u$ now represents the pressure difference between the second phase and the first phase (after and before injecting the currents $\epsilon_i$ at output nodes). Indeed, denoting $v^{\rm free}$ and $v^{\rm nudged}$ the node pressures at equilibrium in the first phase and second phase, respectively, we have
\begin{align}
    \label{eq:link-1}
    L \cdot v^{\rm free} & = I, \\
    \label{eq:link-2}
    L \cdot v^{\rm nudged} & = I - E,
\end{align}
where we recall that $E$ is the vector of currents at output nodes.
Eq.~\eqref{eq:link-1} is the same Laplace equation as \eqref{eq:first-phase}, therefore $v^{\rm free} = v$. Moreover, subtracting \eqref{eq:link-2} from \eqref{eq:link-1}, we see that $v^{\rm nudged} - v^{\rm free}$ satisfies the same equation \eqref{eq:second-phase} as the chemical concentration $u$, that is $L \cdot \left( v^{\rm nudged} - v^{\rm free} \right) = - E$. We conclude that $u = v^{\rm nudged} - v^{\rm free}$.
We thus recover the setting of equilibrium propagation~\cite{endtoend2020} with ``free state'' $v^{\rm free}=v$ and ``nudged state'' $v^{\rm nudged}=v+u$. Coupled learning~\cite{Stern2021} is also very closely related, with the difference that nudged states are realized by imposing boundary conditions on the pressures of output nodes, rather than on the currents.

Without loss of generality, let us assume for convenience that $\alpha=-1$. Our learning rule rewritten in terms of the free and nudged states is 
\begin{align}
    \Delta w_{xy} = & - v(x,y)u(x,y) \\
    = & \frac{1}{2}v(x,y)^2 + \frac{1}{2}u(x,y)^2 - \frac{1}{2}\left( v(x,y) + u(x,y) \right)^2 \\
    = & \frac{1}{2} \left( v^{\rm free}(x,y)^2 - v^{\rm nudged}(x,y)^2 \right) + O(\eta^2),
\end{align}
where $\eta$ is the nudge amplitude, i.e. the factor that scales the amplitude of the currents injected in the second phase. Here we have used that $u$ is proportional to the nudge amplitude, i.e. $u(x,y) = O(\eta)$. We have thus recovered the learning rule of equilibrium propagation~\cite{endtoend2020} and coupled learning~\cite{Stern2021} which, at order $1$ in $\eta$, is
\begin{equation}
    \label{eqprop-rule-resistive}
    \Delta w_{xy}^{\rm EP/CL} = k \left( \left( v^{\rm free}(x,y) \right)^2 - \left( v^{\rm nudged}(x,y) \right)^2 \right).
\end{equation} 
In equilibrium propagation and coupled learning, the multiplicative factor in front of the learning rule is $k=\frac{\text{learning rate}}{2 \times \text{nudge amplitude}}$ ; in our setting here, since $\alpha=-1$, the nudge amplitude is the same as the learning rate, therefore $k=\frac{1}{2}$.\\

To understand the discrepancy of $O(\eta^2)$ between our learning rule and the learning rule of EP and CL, let us rewrite $v^{\rm nudged}$ as $v^{\eta}$ to explicitly show the dependence of the nudged state on the nudge amplitude $\eta$. In particular, $v^0 = v^{\rm free} = v$. In our linear flow network, $u$ is a linear response of $\eta$ (specifically $u= - \eta L^{-1} \frac{\partial C}{\partial v}$, see Eq.~\eqref{eq:second-phase}). Combined with the relationship $u = v^{\eta} - v^0$ shown above, this implies that $u = \eta \left. \frac{\partial v^\eta}{\partial \eta} \right|_{\eta=0}$. Therefore:
\begin{align}
    \Delta w_{xy} & = - v(x,y) u(x,y) \\
    & = - \eta \; v(x,y) \left. \frac{\partial v^\eta(x,y)}{ \partial \eta} \right|_{\eta=0} \\
    & = - \frac{\eta}{2} \frac{d}{ d\eta } \bigg| _{\eta =0} \left[ v^\eta(x,y) \right]^2.
\label{eq:chem_sig}
\end{align}
We recover the learning rule $\Delta w_{xy}^{\rm EP/CL}$ of Eq.~\eqref{eqprop-rule-resistive} as a finite difference approximation of Eq.~\eqref{eq:chem_sig} -- hence the term $O(\eta^2)$.

The advantage of our method over EP and CL is that, by using chemical signaling, the system components can now differentiate between ``activation'' and ``feedback'' signals based on their distinct chemical signatures. This happens because the chemical $u$ encodes the same information as $\frac{\partial v^\eta}{\partial \eta}$. Not only this eliminates the need to store information about two separate phases, but also gives a way for exact gradient computation.

On the other hand, one of the strengths of equilibrium propagation and coupled learning is that they can be used for arbitrary physical systems driven by physical equilibration: their learning rule can be written in terms of derivatives of energy with respect to the trainable parameters\cite{Scellier2017,Stern2021,scellier2020deep}, as follows:
\begin{equation}
    \label{eqprop-rule-general}
    \Delta \mathbf{w}^{\rm EP/CL} = 2k \bigg( \dfrac{\partial \mathcal{E}^{\rm free}}{\partial \mathbf{w}} -  \dfrac{\partial \mathcal{E}^{\rm nudged}}{\partial \mathbf{w} } \bigg),
\end{equation}
where $\mathcal{E}^{\rm free}$ is the energy of the system in the free phase and $\mathcal{E}^{\rm nudged}$ is the energy in the nudged phase. For example, Eq~\ref{eqprop-rule-resistive} can be obtained from Eq~\ref{eqprop-rule-general} by taking $\mathcal{E}$ as power dissipation function \cite{endtoend2020,Stern2021}. While the present work focuses on \textit{linear} physical systems, we will leave the study of our algorithm in nonlinear systems for future work.\\

\section{Discussion}

 We present a simple model of a physical learning system that learns via chemical signaling. In our system, the error between the desired behavior and observed behavior at the output nodes is encoded in the form of a feedback chemical signal. These signals travel across the network via diffusion with the weights of the network updating in response to the concentration of the feedback chemical and there is no need for two states, as there is in equilibrium propagation and coupled learning. We also show that this learning rule minimises a cost function via gradient descent even beyond the infinitesimal nudge amplitude or learning rate limit. 

Given the prevalence of backpropagation, we also compare our algorithm to backpropagation. While there are many explicit differences between backpropagation and our algorithm -- mostly because artificial neural networks are very different from physical flow networks -- we can compare the basic ideas between these algorithms. In our model the weight update rule is proportional to two quantities: the ``error term'' $u(x,y)$ which tells how much the pressure drop across the weight $w_{xy}$ must change, and the ``activation term'' $v(x,y)$ which tells the existing pressure drop due to input. This is similar to backpropagation where the weight update is proportional to the presynaptic input and error in the postsynaptic output~\cite{nielsenneural}. However, the difference between these algorithms is seen in the way error information is communicated. In backpropagation, the error at the output layer and the weights projecting onto this output layer determines the error at the penultimate layer and so on~\cite{nielsenneural}. Therefore, the relationship between the error values of two layers only depends on the local weight values connecting them. In our model, once the error at output is known, the error at the neighbouring nodes is determined by the steady state of the system, which means that the relationship between their error depends on all the weights of the network.

While our work here is mostly concerned with \textit{linear} flow networks in the absence of advection (see Appendix A for an exception to this), one can explore how our algorithm can be extended to include it. Moreover, our physical procedure (Sec.  \ref{sec:theory}) may also be applied to nonlinear systems, i.e., systems whose components have nonlinear characteristics. We will leave the mathematical analysis and experimentation of these settings for future work. We also must explore going beyond the quasistatic limit as time scales also constrain biological learning systems. Efforts towards this goal have recently been made {\it in silico} and in experiments using coupled learning~\cite{Stern2021b}. Similar extensions can be implemented using our algorithm.

Nature may indeed be using similar signaling mechanisms that we have elaborated on here. Cells use biochemical signals to structure themselves in response to external conditions to optimize their functionality and, thus, identity~\cite{cellcognition}. Slime mold, a tubular network-like single cell organism uses chemical signals to modify its tube radii in response to food as the external stimuli~\cite{Alim2017}. While the particular chemical still remains unknown, a recently proposed candidate is ATP~\cite{Alim2021}. ATP is crucial for the activity of myosin and, hence, the contractility of the actin cytoskeleton.  An increase in ATP would presumably enhance contractility. One might conclude that enhanced contractility leads to stiffer cells as is found in cells adherent to a substrate. However, for cells in suspension, inhibition of myosin leads to stiffer cells due to accelerated depolymerization of actin filaments~\cite{Chan2015}.
Perhaps suspended cell behavior is more relevant for the tubular structures in slime model than adherent cells behavior. On other hand, given such understanding, we expect similar artificial versions of this signaling mechanism could be used for experimental realizations of our model with its aforementioned extensions.  For instance, an experimental system that makes use of different mechanisms, such as different catalytic reactions on a flexible sheet to drive shape changes in the presence of flows (see Appendix A), is a distinct possibility~\cite{Laskar2022}. Alternatively, perhaps one may implement similar catalytic reactions in stiff sheets that fold. Incidentally, supervised learning in stiff sheets, i.e. origami, has been explored~\cite{Stern2020supervised}.

As for multi-cellular organisms, the brain presumably fine tunes the synaptic strengths of billions of neurons to generate an optimal behavior. For this to happen, feedback signals should not only carry precise credit information to individual neurons but while doing so, they must not interfere with the activation/feedforward signals~\cite{Lillicrap2020}. How the brain does this is still unknown and many models have been developed to explain this phenomenon. For instance, some models invoke the use of error neurons~\cite{schwartz1993computational}, while others assume a temporal segregation of activation and feedback phases~\cite{Stern2021}. Others propose a compartmentalization of individual neurons to spatially separate information as opposed to temporal segregation~\cite{Ording2001}. Our learning mechanism is similar to this last one. It avoids multiple phases by modifying a node to store two kinds of information - an activation signal and a feedback signal. The reason they do not interfere is because the system components can identify them by their chemical signatures - a ubiquitous phenomenon in nature. We have shown how the same network structure used to send an activation signal, can be used to send precise gradient information to individual weights. Therefore, our model optimises a cost function via gradient descent, something that deep neural networks already do to achieve human like functionality~\cite{imagerecog,alphago,languagetrans,speechrecog}. In light of the reasons stated above, we think our model may ultimately help neuroscientists understand credit assignment mechanisms in the brain.

\acknowledgements{The authors thank Sid Mishra, Sam Dillavou, Doug Durian, Andrea Liu, and Nachi Stern for discussion. JMS acknowledges funding support from NSF-DMR-1832002 and an Isaac Newton Award from the DoD.}

\bibliography{biblio.bib} 
\bibliographystyle{ieeetr}

\appendix 

\section{Flow version of the chemical signaling algorithm}
Here we present a model where the chemical signal spreads not via diffusion but via advection. To begin, the pressure at a node depends on the resistance to flow downstream. Therefore, to alter the pressure at a node, we must change the conductance of pipes downstream. Let us say that there is an output node pressure we wish to decrease. We will simply release a chemical at that node which gets carried by the current downstream. This chemical is such that, when it is flowing through a pipe, it increases the conductance of the pipe (e.g. making it thicker). This increase in conductance decreases the resistance to flow which in turn decreases the pressure at the output node. Similarly, when we wish to increase the output node pressure, we must release a different kind of chemical which decreases the conductance of the pipes (e.g. making it thinner). Using this we can tune the network to output desired voltages.
In fact, we observe numerically system optimises a cost function-but not necessarily via gradient descent.

To implement the above idea as a tuning process, consider  
\begin{enumerate}
    \item The input pressures $\{p_i\}$ is applied at output nodes. A supervisor checks the output pressures $\{v\}$ at output nodes and compares them to desired output pressures $\{{v}_{d}\}$ . 
    \item There are two kinds of chemicals, $s_{+}$ and $s_{-}$. $s_{+}$ increases the conductance of the pipe when it passes through it, and vice versa for $s_{-}$. We assume that the output nodes release a chemical whose amount is proportional to the difference between the present output pressures and desired output pressures. At $t=0$ for some output node $a , ~ v(a)  \neq {v}_{d}(a) $, then \begin{align}
        & ~ ~if ~ v(a) > {v}_{d}(a) \implies s_{+}(a) = \lambda(v(a) - {v}_{d}(a))\,\,\, {\rm at}\,\,\, t=0, \\
        & elif ~ v(a) < {v}_{d}(a) \implies s_{-}(a) = \lambda ({v}_{d}(a)-v(a))\,\,\, {\rm at}\,\,\, t=0,
    \end{align}
    Where $\lambda$ is the factor that controls the chemical response given by the node to the difference in pressures. Moreover, $s_{+}(a)$ denotes amount of chemical (e.g. no of molecules) at node``a" .
    \item This chemical is carried by the current in the network. Therefore, in the next time step the chemical flows to the neighbouring nodes of $a$ that are downstream to $a$\footnote{For simplicity we assume that the chemical has negligible diffusion and the only way it can spread is via the network currents}. We call all such downstream neighbours of $a$ as $\mathcal{D}(a)$. Then for all $b \in \mathcal{D}(a): $\begin{equation}
         s_{+}(b,t+1) = s_{+}(a,t) \times \dfrac{i(b,a)}{  \sum_{x \in \mathcal{D}(a) }i(x,a) } \end{equation}
         \begin{equation*}+  {\rm (incoming~chemical~from~other~ nodes)}
    \end{equation*}
    where $i(x,a)$ represents the current from $a$ to $x$. Note that all the chemical initially present at $a$ flows downstream after one time step.
    \item Using the above equation, an N $\times$ N array $\hat{S}_{+}$ is generated, where each entry $i,j$ denotes the amount of chemical passing through the pipe $\{i,j\}$ at step $t$ $\xrightarrow{}$ $t+1$. Let $\hat{W}$ denote the conductance matrix of the graph, where each entry $\{i,j\}$ denotes the conductance of that pipe. Then \begin{align}
        & \hat{W}(t+1) = \hat{W}(t) + \xi(\hat{S}_{+} - \hat{S}_{-}),
    \end{align}
    $\xi$ controls the response of the pipe to the passing chemical.
    \item The new potentials are calculated on the interior vertices using $\hat{W}(t+1)$. Again, the supervisor checks if $v(a) = v_{d}(a)$. The chemical takes some time to reach the boundary nodes, where it drains out of the network~\footnote{The current flows into and out of the network through boundary nodes}. Therefore, the total change in potential due to the chemical released at the output nodes is observed after some amount of time. Therefore, we introduce a time delay $\tau$ before releasing the chemical once again~\footnote{This helps the potentials at the output node to converge nicely at the desired potentials. If this delay is not introduced, the output potential oscillates about the desired potential. Moreover, the delay helps in avoiding the buildup of excess chemical in the network}.
    \item This process is repeated iteratively.
    
    \end{enumerate}

\end{document}